\documentclass[a4paper]{jpconf}
\usepackage{graphicx}
\begin{document}
\title{How beaming of gravitational waves compares to the beaming of electromagnetic waves: impacts to gravitational wave detection}

\author{Andrew L. Miller$^1$ and Thulsi Wickramasinghe$^1$}

\address{$^1$ Department of Physics, 
The College of New Jersey, 2000 Pennington Road, Ewing, NJ 08628, USA}

\ead{millea12@tcnj.edu}

\begin{abstract}
We focus on understanding the beaming of gravitational radiation from gamma ray bursts (GRBs) by approximating GRBs as linearly accelerated point masses. For accelerated point masses, it is known that gravitational radiation is beamed isotropicly at high speeds, and beamed along the polar axis at low speeds. Aside from this knowledge, there has been very little work done on beaming of gravitational radiation from GRBs, and the impact beaming could have on gravitational wave (GW) detection. We determine the following: (1) the observation angle at which the most power is emitted as a function of speed, (2) the maximum ratio of power radiated away as a function of speed, and (3) the angular distribution of power ratios at relativistic and non-relativistic speeds. Additionally the dependence of the beaming of GW radiation on speed is essentially the opposite of the beaming of electromagnetic (EM) radiation from GRBs. So we investigate why this is the case by calculating the angular EM radiation distribution from a linear electric quadrupole, and compare this distribution to the angular gravitational radiation distribution from a GRB.
\end{abstract}

\section{Introduction}
Binary neutron stars, binary black holes and neutron star-black hole pairs are the most likely source of gravitational waves. Emissions from these compact binaries are expected to be short gamma ray bursts (GRBs), which are very short bursts of large amounts of electromagnetic energy. GRBs are divided into two types based on the signal's duration and spectral hardness: ``short'' and ``long'' GRBs$^1$. The origin of ``short'' GRBs is thought to be from compact binaries coalescence (CBC). If gravitational waves are detected by advanced LIGO from CBCs, we can further distinguish between the signals produced by different binary systems. ``Long'' GRBs mostly originate from stellar collapses and are not that relevant here. An overly simplistic equation used to compute the so-called ``isotropic'' energy due to a GRB is given as$^2$:

\begin{equation}
E_{iso} = \frac{4\pi(BC){D_{L}}^{2}f}{1+z}
\label{e_iso}
\end{equation}

where $E_{iso}$ refers to the energy due to the GRB that is emitted in all directions equally, $z$ is the redshift, $D_L$ is the luminosity distance to the source, $f$ is the fluence measured in the High Energy Transient Explorer (HETE-2) waveband, and BC is the bolometric correction for this waveband for GRBs. $E_{iso}$ is taken to be ~$10^{-2} M_{sun}c^2$. However it is not certain that the CBC sources in LIGO's sensitivity range will emit GRBs isotropically. In fact if the bursts are at lower speeds, they will be beamed along the polar axis, which might present a better chance of detection. However it has been determined that if the GWs are beamed along the polar axis, the power due to these signals would have been underestimated by the factor $u$:

\begin{equation}
u=1+\frac{11x}{16}+\frac{11x^2}{16}+\frac{x^3}{16}+\frac{x^4}{16}
\end{equation}
where $x$ = $\cos\psi$ and $\psi$ is the opening angle around the normal into which the bursts are beamed$^3$.

And if GRBs are emitted isotropicly, they might not be strong enough to be detected, since $E_{iso}$ varies among order of magnitudes for different known GRBs$^4$. So we need to determine what proportion of this variable amount of power is actually detectable.

Additionally, we know that gravitational waves are the result of a time-varying quadrupole moment of a source, but there exists a fundamental difference in the way that electromagnetic radiation from an accelerated point charge behaves at relativistic speeds compared to gravitational radiation from an accelerated point mass at relativistic speeds:

\begin{equation}
\frac{
dE}{d\Omega d\omega} = \frac{e^2}{4\pi^2 c}\beta^2 \gamma^2\frac{\sin^2\theta}{(1-\beta\cos\theta)^2}
\label{em_beam}
\end{equation}

\begin{equation}
\frac{dE}{d\Omega d\omega} = \frac{Gm^2}{4\pi c^2}\beta^4\gamma^2\frac{\sin^4\theta}{(1-\beta\cos\theta)^2}
\label{gw_beam}
\end{equation}

Here $\frac{dE}{d\Omega d\omega}$ is the energy emitted per solid angle $\Omega$ per angular frequency $\omega$, $\theta$ is the observation angle, and the rest of the variables carry their usual meanings. At relativistic speeds, equations \ref{em_beam} and \ref{gw_beam} indicate that EM radiation is beamed and that GW radiation is not (it is in fact isotropic). We investigate why this fundamental difference occurs by accelerating a linear electric quadrupole (two dipoles back-to-back) and calculating its power distribution. There has been work done on calculating the electric and magnetic fields of a dipole in motion, and the corresponding power distributions as a function of $\theta$, the observation angle$^5$. In this case, the power distribution of a linearly accelerating dipole is even more beamed than for an accelerated point charge. However, no such calculation has been done for a linear electric quadrupole in motion, nor has an angular power distribution ever been computed.

\section{Methods}

Even though GRBs come from in-spiraling binaries, we approximate GRBs as accelerated point masses, whose gravitational power distribution is given by equation \ref{gw_beam}. We ask the following questions: (1) at what observation angle do we receive the greatest ratio of power, (2) how is this ratio of power affected by speed of the source, and (3) given a relativistic speed, what will be the source's angular distribution of power?

In order to answer question (1), we differentiate equation \ref{gw_beam} with respect to $\theta$, and determine the angle that corresponds to a maximum in emitted power. We vary the speed $\beta$ to see how the maximum angle changes as the source moves more and more relativistically. For question (2), we integrate equation \ref{gw_beam} with respect to $\theta$ to determine the total power emitted over all solid angles. After, we compute the ratio of maximum power to total power for a given source and vary the speed to determine the impact of changing speeds on the ratio of power emitted. For question (3), we plot the ratio of power at a given speed to the total power (integrated over all angles) as a function of the angle $\theta$.

\begin{figure}[ht!]
\centering
\includegraphics[width=0.3\columnwidth]{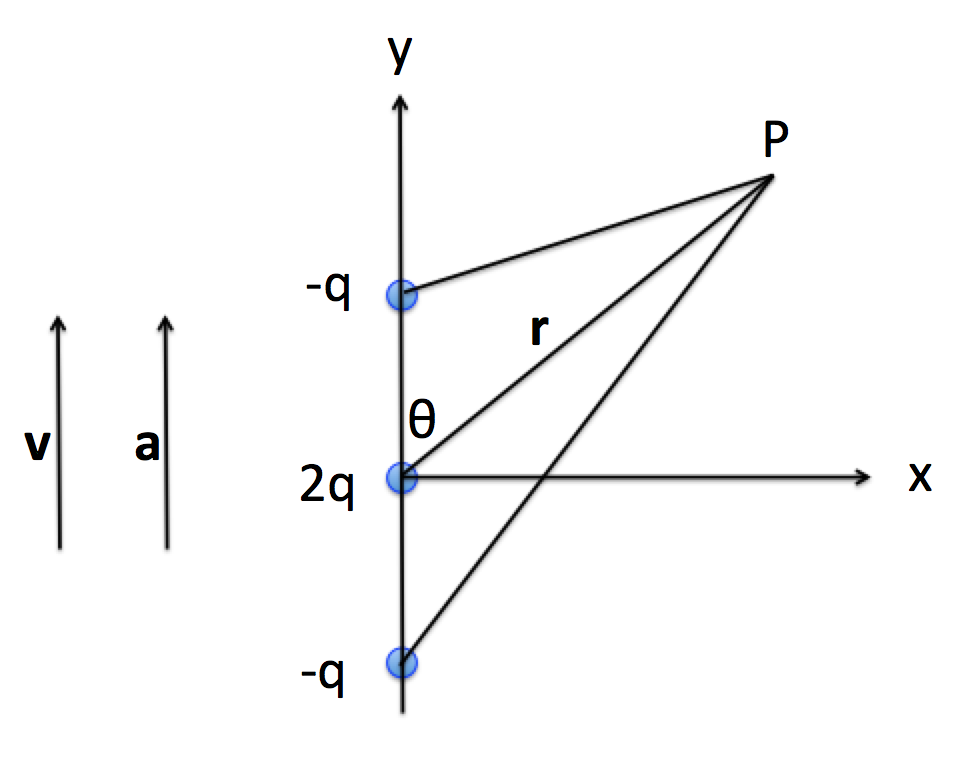}
\caption{Model of accelerating linear electric quadrupole.}
\label{lin_elec}
\end{figure}

To address the fundamental difference between gravitational and electromagnetic power angular distributions, we find the angular power distribution due to a accelerated electric quadrupole, and see if that is similar to the angular power distribution given in equation \ref{gw_beam}. For simplicity, we accelerate a linear electric quadrupole in one direction, as shown in figure \ref{lin_elec}.

We compute the angular power distribution at point $P$ at a distance $r$ from the center of the linear electric quadrupole, by (1) calculating the retarded scalar and vector potentials V and \textbf{A} from the Lienard-Wiechert potentials, (2) finding \textbf{E} and \textbf{B} (keep only 1/r terms), (3) computing the Poynting vector \textbf{S}, and (4) finding the angular power distribution from \textbf{S}, with $\frac{dP}{d\Omega}=|\textbf{S}|r^2$.

For simplicity, we assume that the distances from all the charges to the point $P$ are equal to $r$, and that the angles that the negative charges make with the normal are also $\theta$. In doing this, if we see similar angular power distributions between an accelerated linear electric quadrupole and a linearly accelerated point mass, then we can conclude that the ``changing quadrupole'' nature of the GW may be what causes this angular distribution of power.

\section{Results} 
\begin{figure}[ht!]
\centering
\includegraphics[width=0.6\columnwidth]{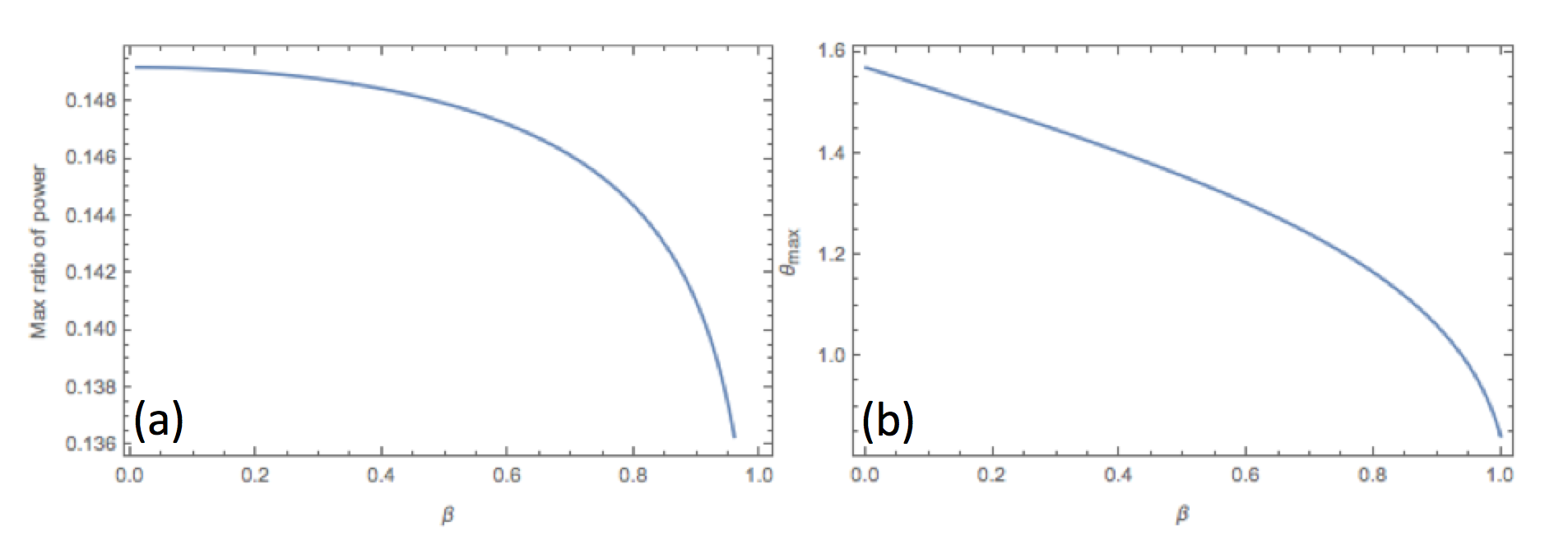}
\caption{(a) Ratio of power emitted at max angle vs. speed, for linearly accelerated point mass, (b) Angle at which max power is emitted vs. speed.}
\label{pow_vs_spd}
\end{figure}

\begin{figure}[ht!]
\centering
\includegraphics[width=0.4\columnwidth]{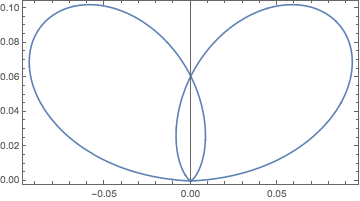}
\caption{Angular power distribution for $\beta=0.999$ for linearly accelerated point mass.}
\label{angle}
\end{figure}

We determine the observation angle per unit solid angle that corresponds to the greatest ratio of power emitted as a function of speed for an accelerated point mass, which is shown in figure \ref{pow_vs_spd}a. Figure \ref{pow_vs_spd}b shows how the ratio of power corresponding to the maximum observation angle changes as a function of speed. We also plot the angular distribution of power for a given speed, to show where we need to look to see the most power.  We see symmetric maxima at $\theta\approx 48^{\circ}$ and a local maximum at $\theta={\pi}/{2}$. See figure \ref{angle}.

We calculated the leading term of the angular power distribution for an accelerated linear electric quadrupole to be:

\begin{equation}
\frac{dP}{d\Omega} \propto a^2 q^4 \frac{\beta^2\cos^2 \theta}{(1-\beta^2\cos^2\theta)^6}
\label{power}
\end{equation}



\begin{figure}[ht!]
\centering
\includegraphics[width=0.4\columnwidth]{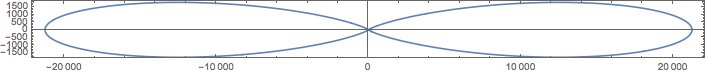}
\caption{Accelerated electric quadrupole power distribution, leading order term, $\beta=0.9$.}
\label{leading_order}
\end{figure}

\begin{figure}[ht!]
\centering
\includegraphics[width=0.4\columnwidth]{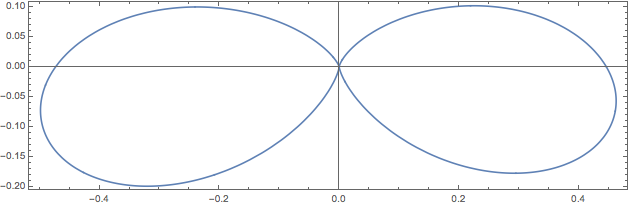}
\caption{Accelerated electric quadrupole power distribution, $\beta=0.1$.}
\label{beta_0_1}
\end{figure}


There are other terms in equation \ref{power} with higher-orders of $\beta$ and $\cos\theta$. However, when neglected, they do not change the angular behavior at high speeds. See figure \ref{leading_order}. At lower speeds, the distribution is relatively isotropic, as shown in figure \ref{beta_0_1}, and the lower-order terms cause a downward ``bending'' of the isotropic distribution.

\section{Discussion}

Based on figure \ref{pow_vs_spd}, only a small fraction (12\%-15\%) of power from a GRB per unit solid angle is received by us even if we are observing it from the optimal angle. Additionally a GRB emitted directly at us at $\beta=0.999$ will not yield the greatest ratio of power ($\theta_{max}\approx48^{\circ}$). We have also seen that the angle at which we will see the most power decreases at higher speeds. Gravitational wave detection is hindered by the angular distribution of accelerated point masses: we therefore believe that radiation from bursts will be harder to see in most scenarios. Even in the best case scenario (at the optimal viewing angle), only around 15\% of the total power per unit solid angle will reach us. The event will need to be very strong and emitted at the optimal angle to be seen.

Figures \ref{leading_order} and \ref{beta_0_1} show that the quadrupole moment is probably not responsible for the differing EM and GW angular power distributions. This is because the EM radiation due to an accelerating electric quadrupole at relativistic speeds is beamed along the observation axis, and the radiation is isotropic at lower speeds. Gravitational radiation from an accelerated point mass behaves exactly opposite to this.

We recognize that a better quadrupole to use would have been a square of charge in two dimensions centered at the origin, since it does not have the higher-order multipoles that a linear electric quadrupole does. We think this calculation would yield more insight into the reasons why changing quadrupole moments drive gravitational wave emission.

\section{References}

\end{document}